\documentclass[structabstract]{aa}  
\usepackage{graphicx}
\usepackage{txfonts}
\begin{document}
   \title{Testing circumstellar disk lifetimes in young embedded clusters
             associated with the Vela Molecular Ridge\thanks{Based
               on observations collected at the European Southern
               Observatory, Cerro Paranal, Chile, programme 074.C--0630}}

%  \subtitle{I. Overviewing the $\kappa$-mechanism}

   \author{F. Massi   
          \inst{1}
          \and
          E. Di Carlo\inst{2}
	  \and
          C. Codella
          \inst{1}
          \and
          L. Testi
          \inst{1,3}
          \and
          L. Vanzi
          \inst{4}
	  \and
          J.\ I. Gomes
          \inst{5,6}
%\fnmsep\thanks{Just to show the usage of the elements in the author field}
          }

   \institute{ INAF - Osservatorio Astrofisico di Arcetri,
              Largo E.\ Fermi 5, I--50125 Firenze, Italy\\ 
              \email{fmassi,codella@arcetri.astro.it}
         \and
             INAF - Osservatorio Astronomico di Collurania-Teramo, 
             Via M.\ Maggini, I--64100 Teramo, Italy\\
             \email{dicarlo@oa-teramo.inaf.it}
        %     \thanks{The university of heaven temporarily does not
        %             accept e-mails}
         \and
             European Southern Observatory, Karl Schwarzschild str.\ 2,
             D--85748 Garching, Germany \\
             \email{ltesti@eso.org}
         \and
             Departamento de Ingeniera Electrica, Pontificia Universidad
	     Catolica de Chile, Av.\ Vicu\~{n}a Mackenna 4860,
             Santiago, Chile \\ 
             \email{lvanzi@ing.puc.cl}
         \and
             Centro de Astronomia e Astrofisica da Universidade de Lisboa,
             Tapada da Ajuda, 1349-018 Lisboa, Portugal
         \and
             Centre for Astrophysics Research, University of Hertfordshire,
             College Lane, Hatfield, Hertfordshire, AL10 9AB, UK\\
             \email{j.gomes@herts.ac.uk}
             }

   \date{Received; accepted}

% \abstract{}{}{}{}{} 
% 5 {} token are mandatory
 
  \abstract
   {The Vela Molecular Ridge hosts a number of young embedded star clusters 
   in the same evolutionary stage.}
   {The main aim of the present work is testing whether the fraction of members
    with a circumstellar disk in a sample of clusters in the cloud D of the
    Vela Molecular Ridge, is consistent
    with relations derived for larger samples of star clusters with an age spread. 
    Besides, we   
    want to constrain the age of the young embedded star clusters associated
    with cloud D.}
   {We carried out $L$ ($3.78$ $\mu$m) photometry on images of six young embedded star
    clusters associated with cloud D of the Vela Molecular Ridge, taken with ISAAC at
    the VLT. These data are complemented with the available $HK_{\rm s}$ photometry.
    The 6 clusters are roughly of the same size and appear to be in the same
    evolutionary stage.
    The fraction of stars with a circumstellar disk was measured in each cluster
    by counting the fraction of sources displaying a NIR excess in 
    colour-colour ($HK_{\rm s}L$) diagrams.}
   {The $L$ photometry allowed us to identify the NIR counterparts of
    the IRAS sources associated with the clusters.
   The fraction of stars with a circumstellar disk appears to be constant within
   errors for the 6 clusters. There is a hint that this is lower for the most
   massive stars. The age of the clusters is constrained to $\sim 1-2$ 
   Myr.}
   {The fraction of stars with a circumstellar disk in the observed sample 
    is consistent with the relations derived from larger samples of
    star clusters and with other age estimates for cloud D.
   The fraction may be lower for the most massive stars.
   Our results agree with a scenario where all intermediate and low-mass stars
   form with a disk, whose lifetime is shorter for higher mass stars.}


   \maketitle
%
%________________________________________________________________

\section{Introduction}

Open clusters have long been considered as suitable laboratories to test
stellar evolution theories and sample the Initial Mass Function (IMF). More recently, 
Near-Infrared (NIR) cameras unveiled the precursors of open clusters: young embedded
star clusters. Even more interestingly, it is now clear that
a significant fraction of star formation
in Giant Molecular Clouds occurs in young embedded clusters. But less
than $\sim 10$ \% of these emerge from molecular clouds as bound
open clusters (see review by Lada \& Lada \cite{la:la}).  
Young embedded clusters are now considered as even more reliable
sites where to measure an IMF, although in this case one has
to take into account the bias introduced by effects such as heavy
reddening and the NIR excess displayed by young pre-main sequence (pms) stars
(see, e. g., review by Scalo \cite{scalo}). 

Giant molecular clouds hosting a number of young embedded clusters
of the same age represent much more valuable laboratories, allowing one
to test even a higher number of physical mechanisms than in single star clusters. 
In this respect, one of the most interesting regions is the Vela
Molecular Ridge (VMR), first studied in the CO(1--0) by Murphy \&
May (\cite{mur:may}). The VMR is a Giant Molecular Cloud complex
located in the galactic plane, in the outer Galaxy ($260\degr \sol l
\sol 272\degr$, $|b| \sol 2\degr$). Murphy \& May (\cite{mur:may})
subdivided the region in 4 main clouds, called A, B, C and D.
The issue of distance was discussed by Liseau et al.\ 
(\cite{liseau}) who concluded that A, C and D lie at $700 \pm 200$ pc.
The molecular gas distribution of the VMR was then studied with a higher
resolution by Yamaguchi et al.\ (\cite{yama}), whereas Elia et al.\
(\cite{elia:07}) and Massi et al.\ (\cite{massi:07}) mapped a $1\degr
\times 1\degr$ sky area towards cloud D
in the CO(1--0) and $^{13}$CO(2--1) transitions and in the continuum 
$1.2$ mm emission, respectively, with the SEST. The dynamical structure
of the molecular gas suggests an age of $\sim 1$ Myr for this part of cloud D.
Star formation in clouds A, C and D 
(both in young embedded clusters and in isolation) 
was studied in a number of works
(to quote but a few: Liseau et al.\ \cite{liseau}; Lorenzetti et al.\ \cite{dloren},
\cite{dloren02}; Wouterloot \& Brand \cite{brandy}; Massi et al.\ 
\cite{massi:99}, \cite{massi:00}, \cite{massi:03}; Giannini et al.\
\cite{gianni05}, \cite{gianni07}; Burkert et al.\ \cite{burk};
Caratti o Garatti et al.\ \cite{CoG};
Baba et al.\ \cite{baba04}, \cite{baba06}; Apai et al.\ \cite{apai};
Thi et al.\ \cite{thi};
De Luca et al.\ \cite{deluca}; Ortiz et al.\ \cite{ortiz};
% 
%  CORREZIONE 2
%
Netterfield et al.\ \cite{netter}; Olmi et al.\ \cite{olmi}).
%
% FINE CORREZIONE 2
%

Massi et al.\ (\cite{massi:06}) exploited the natural
``laboratory'' provided by cloud D
by selecting a sample of 6 small young embedded clusters in a similar
evolutionary stage that they observed in the NIR $JHK_{\rm s}$ bands.
They studied the IMFs and tried to constrain the
age, as well. It could be shown that the IMFs are consistent each others
and consistent with a standard IMF
(e. g., the one proposed by Scalo \cite{scalo}). But the cluster age could only be
loosely constrained to lie in the interval $\sim 1-5$ Myr. Therefore, it is 
critical to obtain a better age determination.

A valuable tool to determine the age of young embedded clusters is
their content of pms stars with a circumstellar disk, that are
identifiable due to their NIR excess with respect to
photospheric emission (Haisch et al.\
\cite{hai:01}, Hillenbrand \cite{hille05}). In principle, this
can be easily achieved by counting cluster members displaying 
a NIR excess and ones just falling in the reddening band of the
main sequence, in a $JHKL$ or $HKL$ colour-colour diagram.  
In the case of the VMR, it is tempting to exploit the large number of young 
embedded clusters of the same age to also test whether the fraction
of stars with a circumstellar disk only depends on the cluster age
or other effects have to be taken into account.
In a scenario where all protostars accrete through a disk, the
fraction of stars with a circumstellar disk in a cluster
must depend only on disk lifetimes, hence it is expected to
be only a function of cluster age, at least in clusters of similar sizes.
With these two aims in mind, we selected a sample of 6 young embedded
clusters associated with the part of cloud D mapped by Elia et al.\
(\cite{elia:07}) and Massi et al.\ (\cite{massi:07}),
and imaged them in the
$L$ ($3.78$ $\mu$m) band. The obtained photometry was then complemented
by $HK_{\rm s}$ photometry already available. 

The paper layout is the following: observations and data reduction
are described in Sect.~\ref{obs:sec}, the results are discussed
in Sect.~\ref{Res:ults} and summarised in Sect.~\ref{Discu:sion}.

%__________________________________________________________________

\section{Observations and data reduction}
\label{obs:sec}

%
%   OSSERVAZIONI L
%
We selected the same fields as Massi et al.\ (\cite{massi:06}), excepted
IRS19 since the NIR counterpart of the IRAS source is too bright in $L$
and would saturate the detector. We replaced it with IRS22, another
site hosting a young embedded star cluster and associated with cloud D.
These are the most populated clusters found in cloud D of the VMR.
%The six fields (see Tab.~\ref{table:fields})
%
%  CORREZIONE 1
%
The six fields are listed in Tab.~\ref{table:fields}, along with 
the gas and stellar masses associated with the stellar clusters.
The total gas mass is inferred from Table~3 of Elia et al.\ (\cite{elia:07}),
whereas the total mass in dense gas from Table~1 of Massi et al.\ 
(\cite{massi:07}). We note that the listed total gas masses are less than the total
masses in dense gas in three cases. The gas mass is derived in
Elia et al.\ (\cite{elia:07}) by using CO(1--0) and $^{13}$CO(2--1) observations
(and assuming LTE conditions)
with the highest spatial resolution obtained so far. Nevertheless, their $^{13}$CO(2--1)
observations are spatially undersampled by a factor 2 and their sensitivity is
low ($\sim 0.7$ K), meaning that the computed masses have to be considered as mere lower 
limits. In fact, Yamaguchi et al.\ (\cite{yama}) find much higher gas masses
by using CO(1--0) and $^{13}$CO(1--0) observations with a larger beam. 
%
% FINE CORREZIONE 1
%

The fields
were observed with the camera ISAAC (Moorwood et al.\ \cite{moorw})
at the ESO-VLT telescope (Cerro Paranal, Chile) through the broad-band filter $L$
centred at $3.78$ $\mu$m, in the LW imaging mode. 
The pixel scale is $\sim 0.07$ arcsec/pixel
and the field of view is $73 \times 73$ arcsec$^{2}$.
All observations were made in service mode. IRS 17 and IRS 18 were observed
on 24/01/2005, IRS 20, IRS 21 and IRS 22 were observed on 25/01/2005 and
IRS 16 on 20/02/2005. On each night, images of the standard star HD75223 
(van der Bliek et al.\ \cite{vdb}) and dark frames
were also taken. Sky flats were acquired on 24/01/2005. 

%      FIELD DESCRIPTION                                       
%_____________________________________________________________
%
\begin{table*}
\caption{Description of the observed fields.
\label{table:fields}}     
\centering                          % used for centering table
\begin{tabular}{c c c c c c}        % centered columns (4 columns)
\hline\hline                 % inserts double horizontal lines
IRAS$^a$ 
    & RA(2000) & DEC(2000) & Total gas mass$^b$ & Total mass in dense gas$^{c}$ 
       & Cluster stellar Mass$^{d}$ \\    
source & & & ($M_{\sun}$) & ($M_{\sun}$) & ($M_{\sun}$) \\    
\hline                        % inserts single horizontal line
IRAS08438--4340 (IRS16) & 08:45:35.8 & --43:51:00 & 99 & 105 & 103 \\   
IRAS08448--4343 (IRS17) & 08:46:35.0 & --43:54:30 & 48 &  88 &  90 \\   
IRAS08470--4243 (IRS18) & 08:48:47.7 & --42:54:22 & ($^e$) &  ($^f$)   &  89 \\   
IRAS08476--4306 (IRS20) & 08:49:26.6 & --43:17:13 & 13 &  21 &  70 \\   
IRAS08477--4359 (IRS21) & 08:49:32.9 & --44:10:47 & 219 & 60 &  68 \\   
IRAS08485--4419 (IRS22) & 08:50:20.7 & --44:30:41 & ($^e$) & ($^f$)   & $>38^g$    \\   
\hline                                   %inserts single line
\end{tabular}

\vspace*{1mm} $^a$ also listed the nomenclature used by Liseau et al.\
(\cite{liseau}); $^b$ derived from CO(1--0) and $^{13}$CO(2--1) observations
by Elia et al.\ (\cite{elia:07}); $^{c}$ derived from continuum 1.2 mm
observations by Massi et al.\ (\cite{massi:07}); $^{d}$ estimated by
Massi et al.\ (\cite{massi:06}); $^e$ found CO(1--0) emission
(Liseau et al.\ \cite{liseau}); $^f$ found a dense core at 1.2 mm
centred at the cluster (Massi F., private comm.); $^g$ inferred from
the data in Massi et al.\ (\cite{massi:03}). 
\end{table*}
%
%_____________________________________________________________

For every field, two sets of 12 chopping sequences of images were taken. In each
chopping sequence, the pointing was switched between two different positions 
(hereby called on and off, although both point towards the target) 
30 times, by moving the telescope secondary mirror.
The integration time in each on and off position results from an average of
(NDIT$=$) 9 single exposures of (DIT$=$) $0.11$ s. After each chopping sequence, all ons
were averaged together and the same for all offs, then the telescope
was nodded, following an ABBAABBA ...
sequence. The chop throw was $15 \arcsec$ and the only 
difference between the two sets of 12 chopping sequences is the throw angle.
The first sequence is chopped (in equatorial coordinates) east-west, 
whereas the second one in a direction rotated by 45 degrees with respect
to the previous. A small jitter ($\sim 5 \arcsec$) was used between ABBA
chopping sequences. The nodding was in the same direction as the
chopping, with the same throw. The total integration time per field
is $\sim 24$ minutes. 
 
Data were reduced using IRAF\footnote{IRAF is distributed by the National
 Optical Astronomy Observatories, which are operated by the Association
 of Universities for Research in Astronomy, Inc., under cooperative
 agreement with the National Science Foundation.} 
routines and following the steps
outlined in the ``ISAAC Data Reduction Guide 1.5'', available online.
Each on and off frame was first dark subtracted and then corrected
for non-linearity according to Sect.~7.4.2 of the Data Reduction Guide
(using updated coefficients). 
After flat-fielding, for each single chopping sequence
the averaged off frame was subtracted from the averaged on frame.
All obtained on--off frames were then corrected for bad pixels. 

An on--off frame consists of displaced ``positive'' and ``negative'' images
of stars (due to the chop throw). Each AB cycle yields
four different shifted images of the same field, i. e. two
on--off and two off--on. For each AB cycle,
by subtracting the on--off B from the on--off A one obtains
a new image where residual sky patterns are effectively cancelled.
However, there are now two symmetric ``negative'' stars with respect
to the corresponding positive ones (twice in counts). This image can
be multiplied by $-1$ and shifted to obtain two more independent
images (also with negative stars).
Therefore, we produced 4 sky-subtracted,
shifted images for each AB cycle.  
Finally, all sky-subtracted images of a same field were registered
and averaged together, after removal of the ``negative'' stars,
to yield the final image.  
We also constructed a median final image in the same way, and
an average and a median image for each of the two 
sets with different chopping throw angle.
Note that, because of chopping and nodding, only the very central area
of the final image exploits the full $\sim 24$ minutes of integration.
The outermost image area was exposed for $6$ minutes. The effective 
integration time increases from the edge to the centre. 
%This should
%roughly compensate for the expected increase in extinction from the field
%%edge to the centre.

The seeing was quite good: we found average PSF FWHMs of
$\sim 0.3 \arcsec$ (IRS16), $\sim 0.5 \arcsec$ (IRS17 and IRS18), and
$\sim 0.4 - 0.6 \arcsec$ (IRS20, IRS21 and IRS22).
Photometry on the $L$ images was performed by using DAOPHOT tasks
in IRAF.
For each field, the list of detections obtained with DAOPHOT
was visually checked on the final median images, to search for both false 
and missed detections. 
It was very useful having images with different chopping throw angles to
check for stars that could have been cancelled in the preliminary on--off
subtraction.  
Aperture photometry was carried out with PHOT, by using an aperture radius
$\sim 1$ FWHM and inner and outer radii for the sky annulus of
$\sim 2$ and $\sim 4$ FWHMs. At last, we performed PSF-fitting photometry
with ALLSTAR. 

For IRS16, IRS17, IRS18, IRS20 and IRS21, we complemented our $L$ photometry
with the $JHK_{\rm s}$ photometry (from SofI images taken at NTT/ESO) 
by Massi et al.\ (\cite{massi:06}). We cross-checked
both photometries and redid the $JHK_{\rm s}$ 
photometry on few sources that had gone undetected
by Massi et al.\ (\cite{massi:06}). In the case of IRS17, the group of sources
towards \# 40 of Massi et al.\ (\cite{massi:99}), resolved on the ISAAC
image, are barely resolved on the SofI images. Hence, only for
the group of stars close to \# 40 we adopted the
$JHK_{s}$ photometry (from ISAAC images taken at the VLT) by Giannini et 
al.\ (\cite{gianni05}).

For IRS22, we performed new $HK_{\rm s}$ photometry on SofI images taken
on 31/12/2007 as part of ESO programme 080.C--0836. These used the
Large Field imaging mode (like in Massi et al.\ \cite{massi:06}) with a plate
scale of $\sim 0.29$ arcsec/pixel, yielding a f.o.v. of $4.9 \times
4.9$ arcmin$^{2}$. Both at $H$ and at $K_{\rm s}$, the field was imaged 
in a $2 \times 2$ square grid, chosen in such a way that
all frames overlap on a central 
area $\sim 2 \times 2$ arcmin$^{2}$ containing the cluster. For each of the
4 positions, 10 dithered integrations were performed, each one a mean of
12 exposures of 2 s. After each on-source integration, the telescope was moved
to a location off-source and a sky image taken with the same averaged
exposure time. Dome flat field images were taken before the observations.
Data reduction was carried out by using standard IRAF routines.
All frames were corrected for cross-talk and flat-fielded. For each on-source
frame, a sky frame was constructed by median-combining the 4 closest (in time)
off-source frames after star removal. All on-source frames were then 
sky-subtracted and corrected for bad pixels. At last, they were registered
and combined together through their average. The $L$-field is contained in
the central $\sim 2 \times 2$ arcmin$^{2}$ area of the $HK_{\rm s}$ mosaics,
so the total integration time of interest is 16 minutes.  

Photometry in the $HK_{\rm s}$ band was carried out with DAOPHOT tasks in IRAF.
We searched for stars with DAOFIND and, then, visually inspected the
images to remove false detections and/or add missed detections (after
a cross-check with the $L$ image). We used PHOT to perform aperture
photometry, with an aperture radius of $\sim 1$ FWHM and inner and
outer sky annuli of $\sim 2$ and $\sim 4$ FWHMs (with median as a sky estimator).
The PSF FWHM is $\sim 0.9 \arcsec$. Calibration was obtained by
using standard stars imaged during the night from the list
by Persson et al.\ (\cite{persy}).

%
% SORGENTI JHK senza controparte L!
% 
Finally, we combined the $JHK_{s}$ and $L$ photometry together.
Over the total solid angle imaged by ISAAC towards each cluster, 
the number of $L$ sources with a $K_{\rm s}$ counterpart found
is: 107 (IRS16), 67 (IRS17), 133 (IRS18), 49 (IRS20), 71 (IRS21) and 101 (IRS22). 
The number of $L$ sources 
without a $K_{s}$ counterpart is instead: 13 (IRS22), 2 (IRS16), 1 (IRS17), 1 (IRS18) and 0 
(IRS20 and IRS21). The number of $K_{\rm s}$ sources without an $L$ counterpart is:
34 (IRS16), 65 (IRS17), 87 (IRS18), 109 (IRS20), 48 (IRS21) and 105 (IRS22). If 
$K_{\rm s} - L$ is computed
for the $K_{\rm s}$ sources without an $L$ counterpart by using the $L$ completeness limits 
estimated in Sect.~\ref{compl:sec}, most of these 
objects fall on the right of the main sequence reddening band in a $H - K_{\rm s}$ vs.\ 
$K_{\rm s} - L$ diagram.
I. e., they are consistent with reddened or unreddened stars whose $L$ photospheric emission lies below
the completeness limit (in flux). Some or them might even have a NIR excess and still go undetected 
in $L$. This also means that the available $K_{\rm s}$ photometry is deeper (in terms of sampled 
stellar masses) than our $L$ photometry.

\section{Results}
\label{Res:ults}

%
%LIMITE DI COMPLETEZZA IN L
%
\subsection{Completeness limits}
\label{compl:sec}

 To estimate the completeness limit of our photometry, we first constructed histograms of the 
number of sources vs.\ $L$ for each field, with $L$ binned in $0.5$ mag intervals. 
In all histograms, the number of sources increases with increasing
magnitude, peaks, and then decreases. The peak 
is the signature of the efficiency in finding sources starting to decrease.
Using the task ADDSTAR in DAOPHOT, we carried out experiments by adding small sequences of
artificial stars randomly distributed over each frame. We could always retrieve
more than 80 \% of the artificial stars at the central magnitude 
of each histogram peak. 
Furthermore, the magnitudes measured with aperture photometry (PHOT) 
were within $\sim 0.1$ mag of the actual values around the peak magnitude. 

Hence, we estimated completeness magnitudes of $L_{\rm compl} \sim 14.25$
(IRS16, IRS21 and IRS22), $L_{\rm compl}' \sim 13.75$ (IRS 17 and IRS 18) and
$L_{\rm compl}'' \sim 13.25$ (IRS 20), that is 1--$1.5$ magnitudes below the detection limits
($L \sim 14.5 - 15.7$). The latter ones 
were estimated by examining the error vs.\ magnitude
diagrams and locating the faintest stars in each diagram.
%(i. e., the magnitude of the faintest stars in the histograms). 
IRS18 and IRS22, the most populous clusters, also exhibit a second peak in 
the number of sources vs.\ $L$, located $\sim 2$ mag
below the completeness magnitude, that we interpret as an intrinsic feature of
the cluster population.

Using the assumed distance modulus to Vela-D ($9.225$ mag) and an estimate of the maximum
extinction, we can convert
$L_{\rm compl}$ ($L_{\rm compl}'$, $L_{\rm compl}''$) to a limiting absolute magnitude. 
In turn, this can be converted into a limiting
stellar mass if a maximum age for the cluster members is also assumed.
By using the reddening law by Rieke \& Lebofsky (\cite{rie:le}), we derived 
completeness limits in absolute magnitudes of $L \sim 4.45$ ($\sim 3.95, 3.45$)
for $A_{V} \sim 10$ mag, and $L \sim 2.71$ ($\sim 2.21, 1.71$) for $A_{V} \sim 40$ mag. 
To convert these magnitudes into stellar masses, we made use of pms evolutionary tracks by a number
of different authors. Adopting the tracks by Palla \& Stahler (\cite{ps99}), 
by a rough comparison with the colours given by Koornneef (\cite{Koorn}) and
Bessell \& Brett (\cite{bess:brett}), the values for $A_{V} \sim 40$ mag are consistent  
with $M \sim 0.4$ $M_{\sun}$ ($\sim 0.6, 1$ $M_{\sun}$)
for an age of 1 Myr, and $M < 0.1$ $M_{\sun}$ ($<0.2, \sim 0.2$ $M_{\sun}$) for an age of $0.1$ Myr.
Adopting the tracks by D'Antona \& Mazzitelli (\cite{dm94}), these numbers become 
$M \sim 0.6$ $M_{\sun}$ ($\sim 0.8, 1$ $M_{\sun}$) for an age of 1 Myr, and 
$M \sim 0.2$ $M_{\sun}$ ($\sim 0.3, 0.4$ $M_{\sun}$) for an age of $0.1$ Myr. 

The tracks by Siess et al.\ (\cite{siess}), with the colours
of Kenyon \& Hartmann (\cite{ke:ha}), yield $M \sim 0.2$ $M_{\sun}$ 
($\sim 0.3, 0.7$ $M_{\sun}$) for an age of 1 Myr and 
$M < 0.1$ $M_{\sun}$ ($\sim 0.2, 0.3$ $M_{\sun}$) 
for an age of $0.1$ Myr. At last, the tracks by Baraffe et al.\ (\cite{bal98}) 
yield $M \sim 0.13$ $M_{\sun}$ 
for an age of 2 Myr, if the colours by Kenyon \& Hartmann 
(\cite{ke:ha}) are used. However, these colours appear inconsistent with those 
inferred from Baraffe et al.\ (\cite{bal98}). Using the colours
by Koornneef (\cite{Koorn}), one obtains $M \sim 0.7$ $M_{\sun}$. 
In summary, the $L$ completeness limits correspond to stellar masses in the range $0.1$--$0.7$ 
$M_{\sun}$ ($0.2$--$0.8$, $0.2$--1 $M_{\sun}$),
seen through an extinction of $A_{V} \sim 40$ mag. We note that this takes into account the star 
photometric emission only, so lower mass stars are still detectable if they exhibit
a NIR excess.

\subsection{The observed fields}
\label{fields:sec}

\begin{figure}
   \centering
   \includegraphics[angle=-90,width=9cm]{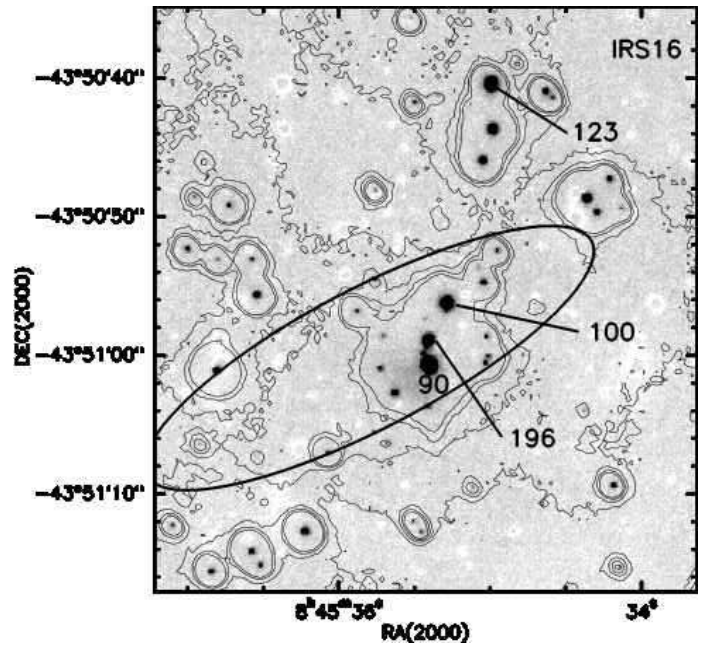}
      \caption{$L$-band image of the inner $\sim 1 \times 1$ arcmin$^{2}$
        region of the cluster associated with IRS16. This is overlaid with
        contours of the emission in the $K_{\rm s}$-band (from the SofI image
        by Massi et al.\ \cite{massi:06}). 
        Contours range from $\sim 10  \sigma$
        of the background in steps of $\sim 10  \sigma$. 
        The IRAS uncertainty ellipse is also drawn,
        and the most prominent NIR sources identified by Massi et al.\
        (\cite{massi:03}) are labelled according to the numbers these authors assigned 
        in their list.
         \label{irs16_field} }
   \end{figure}

\begin{figure}
   \centering
   \includegraphics[angle=-90,width=9cm]{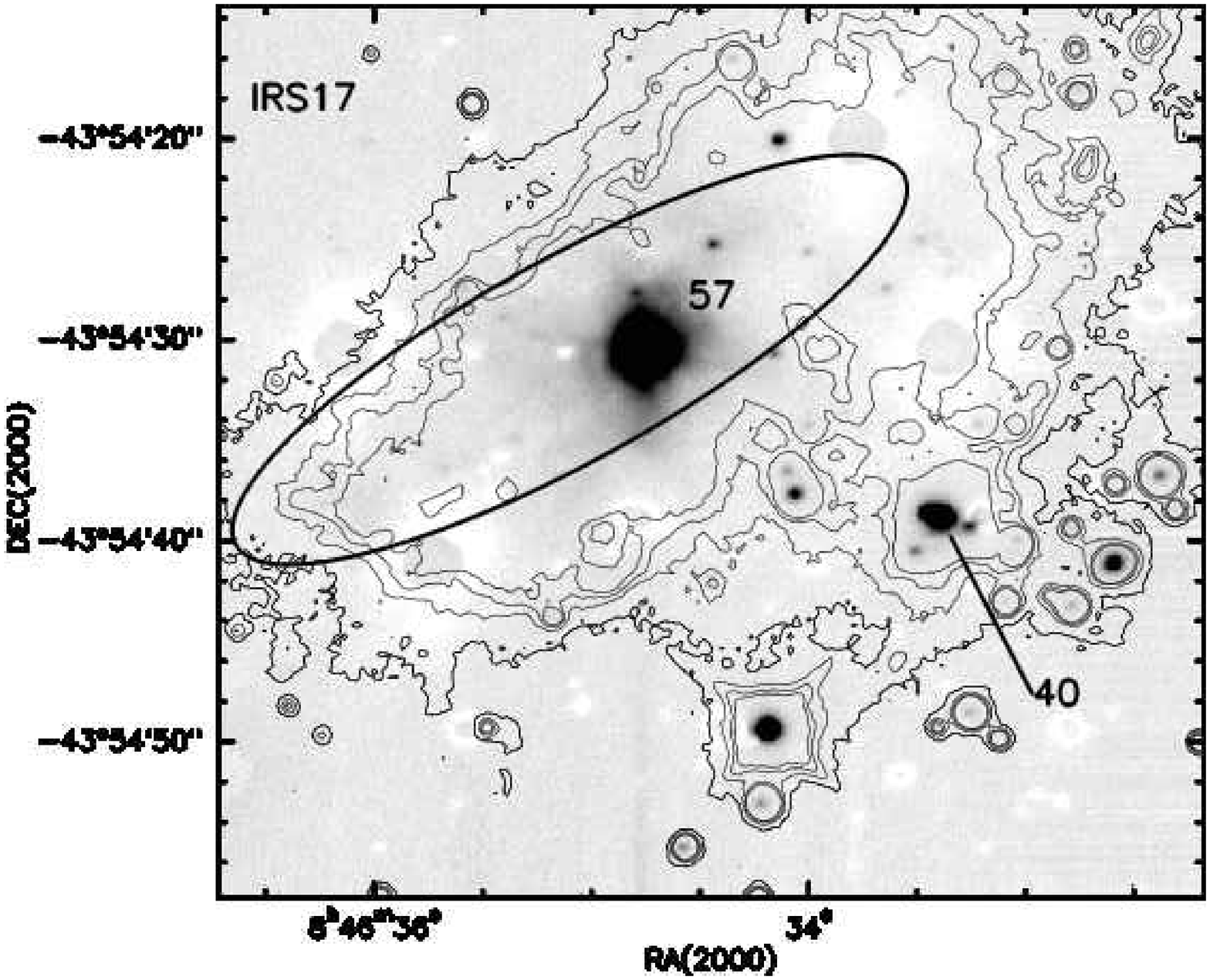}
      \caption{$L$-band image of the inner $\sim 1 \times 1$ arcmin$^{2}$
        region of the cluster associated with IRS17. This is overlaid with
        contours of the emission in the $K_{\rm s}$-band (from the SofI image
        by Massi et al.\ \cite{massi:06}). 
        Contours range from $\sim 10  \sigma$
        of the background in steps of $\sim 10  \sigma$. 
        The IRAS uncertainty ellipse is also drawn,
        and the most prominent NIR sources identified by Massi et al.\
        (\cite{massi:99}) are labelled according to the numbers these authors assigned 
        in their list.
         \label{irs17_field} }
   \end{figure}

\begin{figure}
   \centering
   \includegraphics[angle=-90,width=9cm]{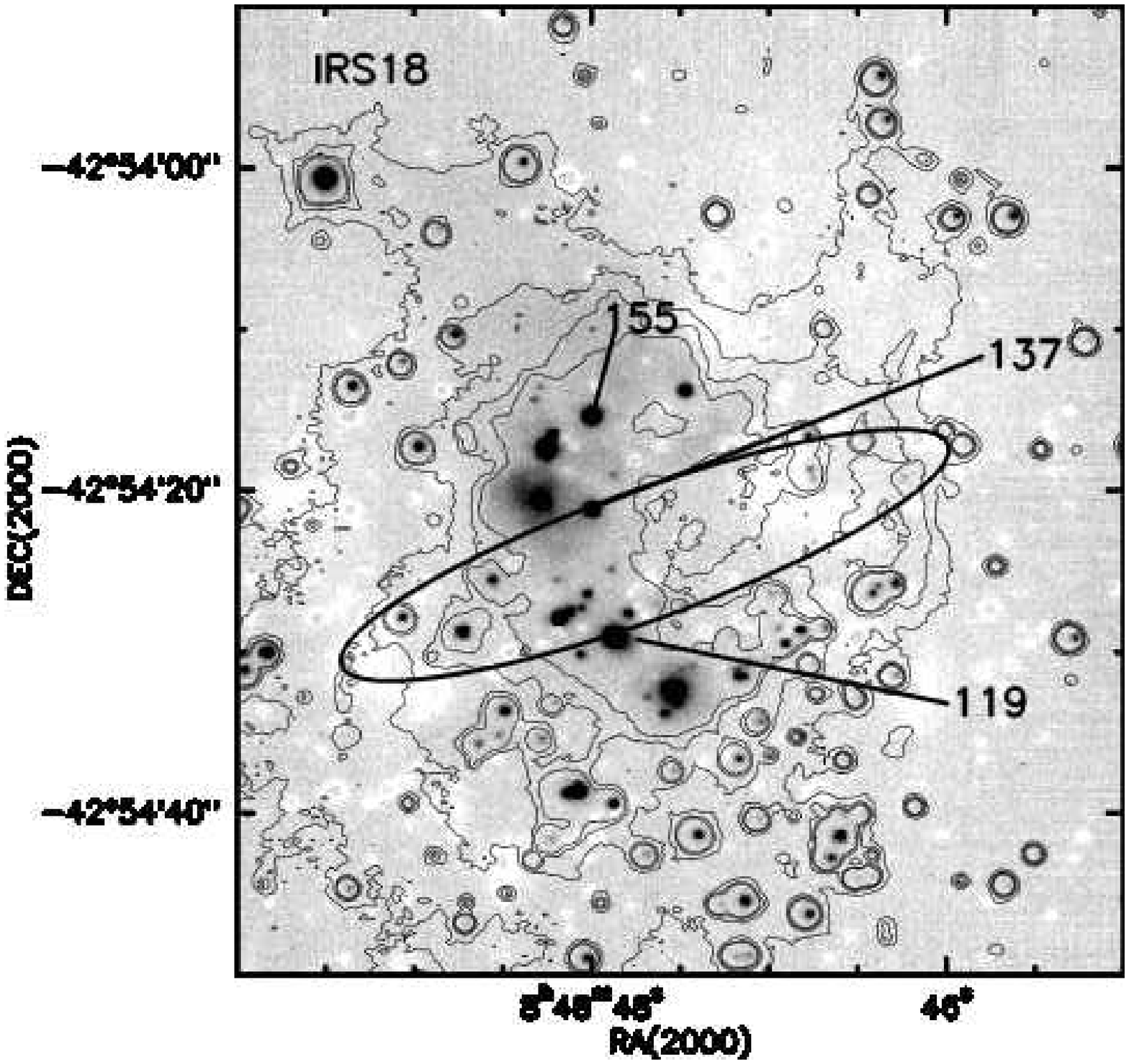}
      \caption{$L$-band image of the inner $\sim 1 \times 1$ arcmin$^{2}$
        region of the cluster associated with IRS18. This is overlaid with
        contours of the emission in the $K_{\rm s}$-band (from the SofI image
        by Massi et al.\ \cite{massi:06}). 
        Contours range from $\sim 10  \sigma$
        of the background in steps of $\sim 10  \sigma$. 
        The IRAS uncertainty ellipse is also drawn,
        and the most prominent NIR sources identified by Massi et al.\
        (\cite{massi:99}) are labelled according to the numbers these authors assigned 
        in their list.
         \label{irs18_field} }
   \end{figure}

\begin{figure}
   \centering
   \includegraphics[angle=-90,width=9cm]{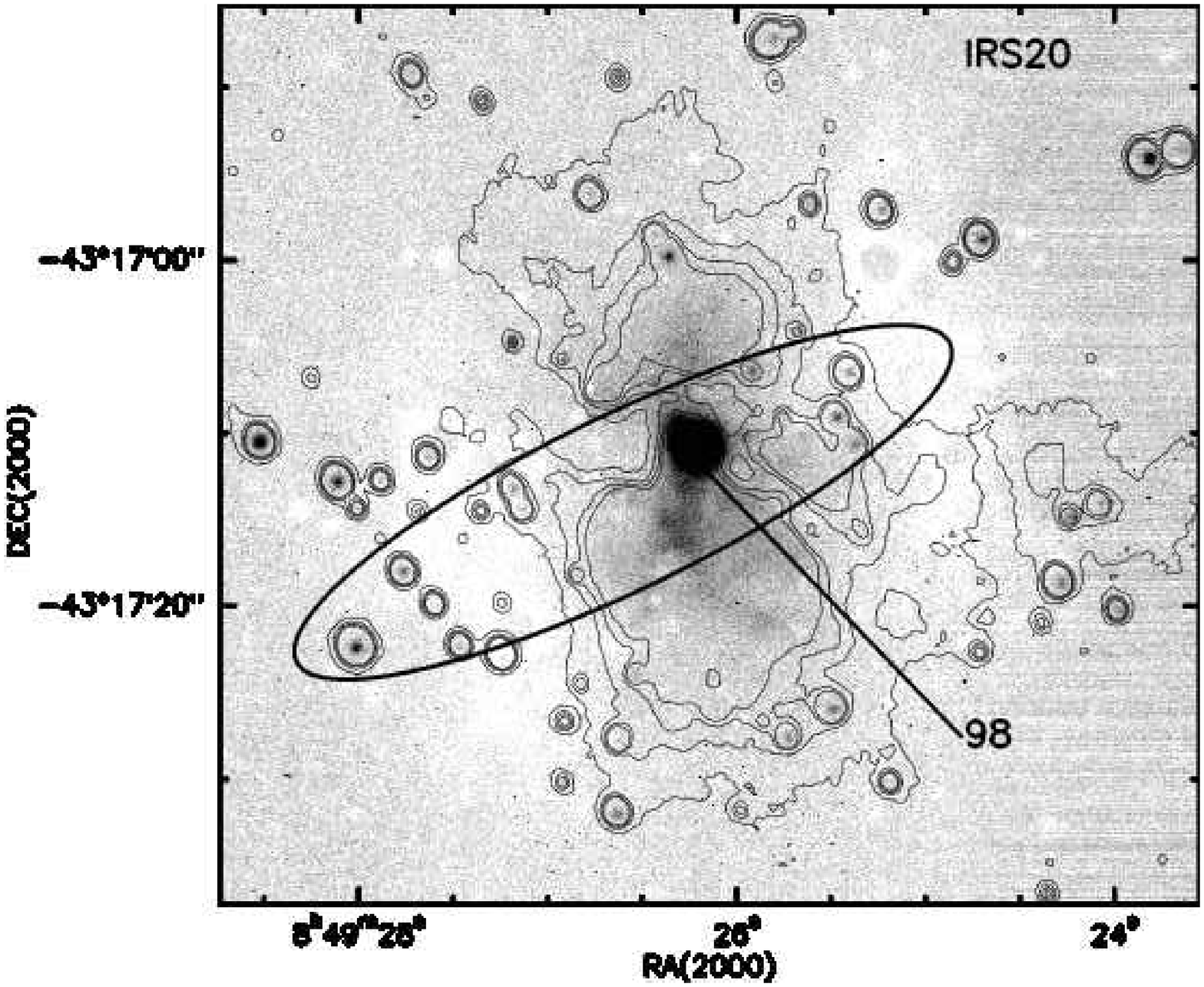}
      \caption{$L$-band image of the inner $\sim 1 \times 1$ arcmin$^{2}$
        region of the cluster associated with IRS20. This is overlaid with
        contours of the emission in the $K_{\rm s}$-band (from the SofI image
        by Massi et al.\ \cite{massi:06}). 
        Contours range from $\sim 10  \sigma$
        of the background in steps of $\sim 10  \sigma$. 
        The IRAS uncertainty ellipse is also drawn,
        and the most prominent NIR sources identified by Massi et al.\
        (\cite{massi:99}) are labelled according to the numbers these authors assigned 
        in their list.
         \label{irs20_field} }
   \end{figure}

\begin{figure}
   \centering
   \includegraphics[angle=-90,width=9cm]{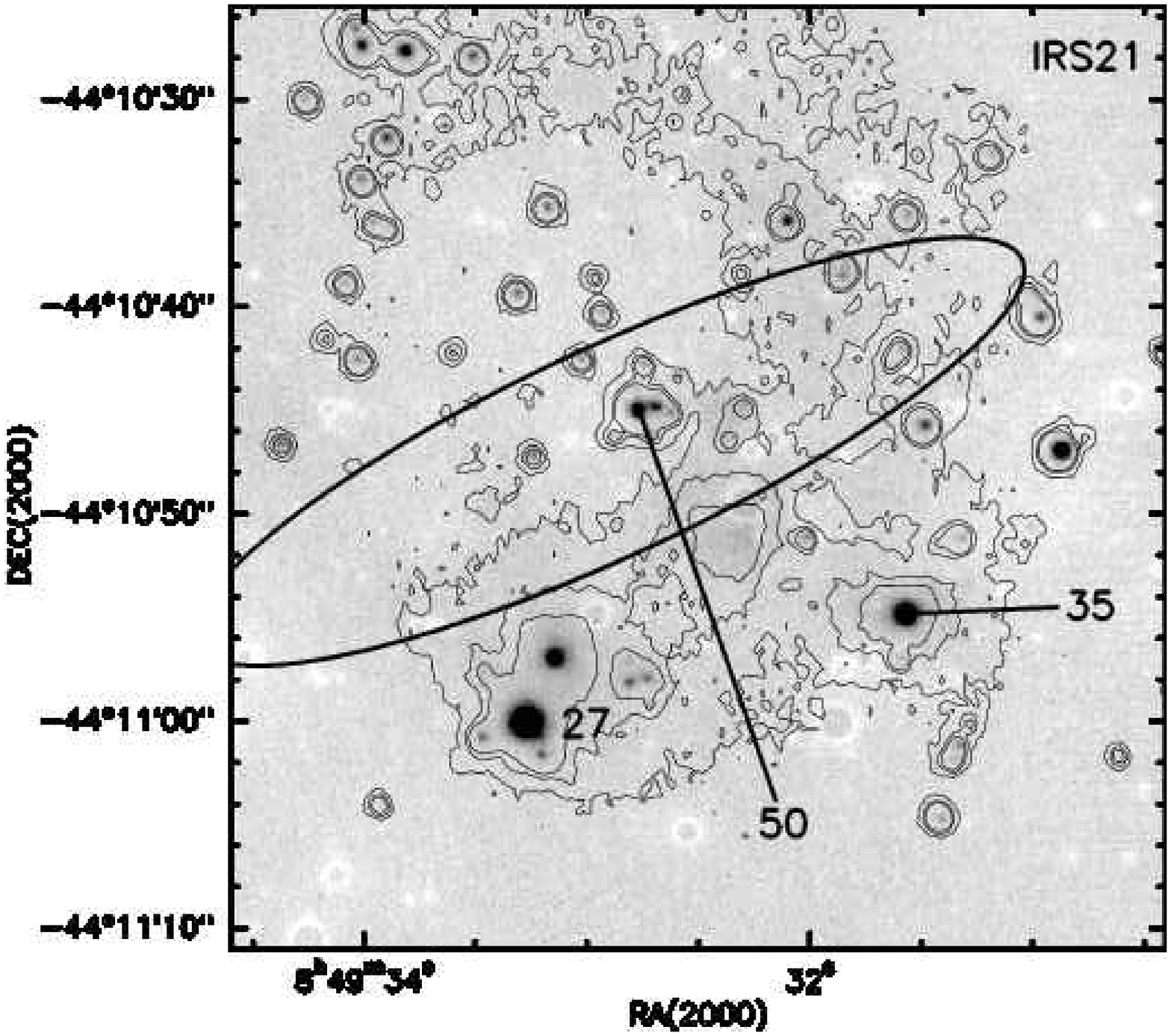}
      \caption{$L$-band image of the inner $\sim 1 \times 1$ arcmin$^{2}$
        region of the cluster associated with IRS21. This is overlaid with
        contours of the emission in the $K_{\rm s}$-band (from the SofI image
        by Massi et al.\ \cite{massi:06}). 
        Contours range from $\sim 10  \sigma$
        of the background in steps of $\sim 10  \sigma$. 
        The IRAS uncertainty ellipse is also drawn,
        and the most prominent NIR sources identified by Massi et al.\
        (\cite{massi:99}) are labelled according to the numbers these authors assigned 
        in their list.
         \label{irs21_field} }
   \end{figure}

\begin{figure}
   \centering
   \includegraphics[angle=-90,width=9cm]{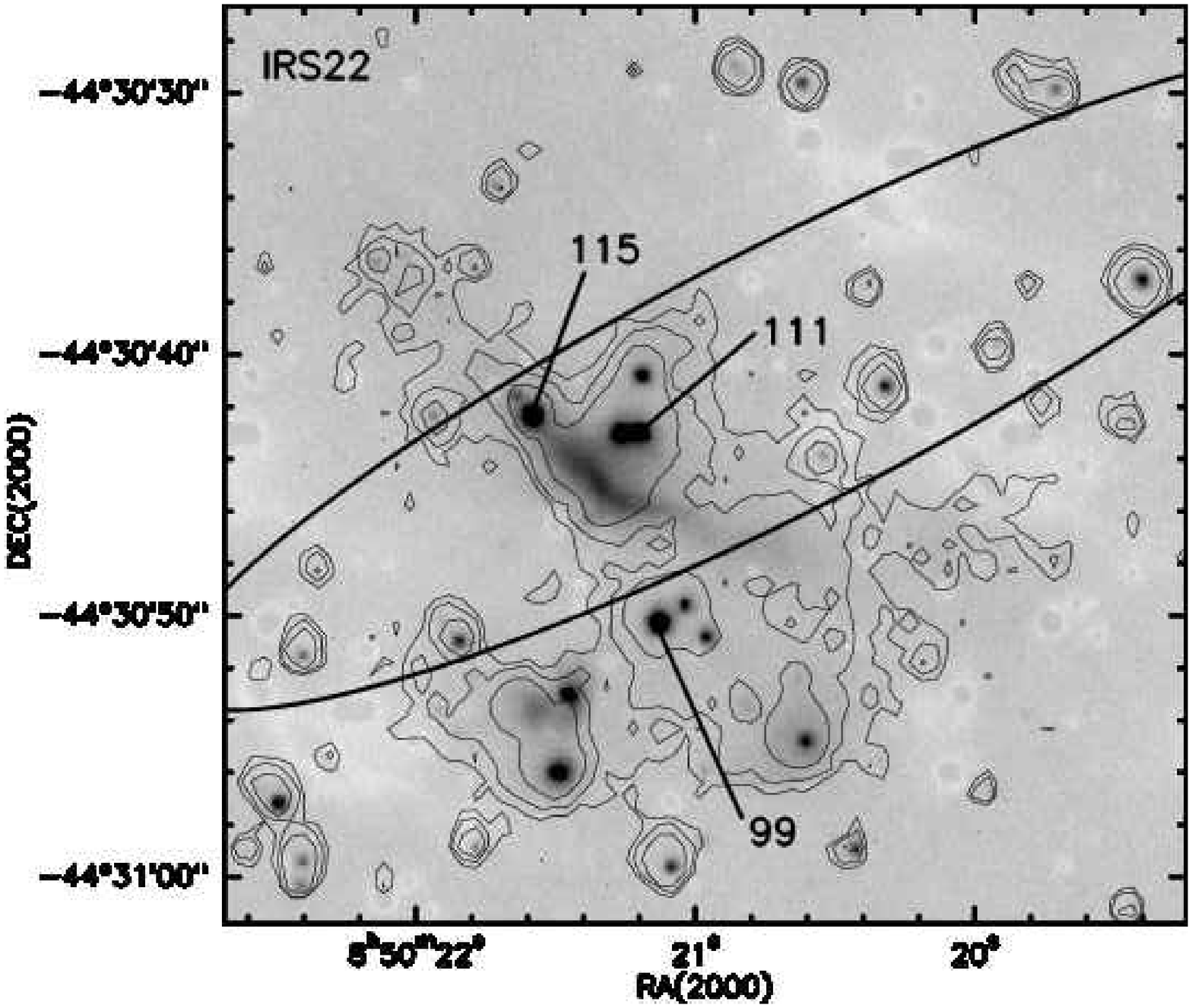}
      \caption{$L$-band image of the inner $\sim 1 \times 1$ arcmin$^{2}$
        region of the cluster associated with IRS22. This is overlaid with
        contours of the emission in the $K_{\rm s}$-band (from the SofI image
        taken in 2008). 
        Contours range from $\sim 10  \sigma$
        of the background in steps of $\sim 5  \sigma$. 
        The IRAS uncertainty ellipse is also drawn,
        and the most prominent NIR sources identified by Massi et al.\
        (\cite{massi:03}) are labelled according to the numbers these authors assigned 
        in their list.
         \label{irs22_field} }
   \end{figure}

Zoom-ins of the inner $\sim 1 \times 1$ arcmin$^{2}$ fields imaged in the $L$-band 
are shown in Figs.~\ref{irs16_field} to \ref{irs22_field}. The $L$ images are
overlaid with contours of the lowest levels of the emission in $K_{\rm s}$ 
(i. e., starting from $\sim 10 \sigma$ of the background) measured on SofI 
(IRS16, IRS17, IRS18, IRS20, IRS21; Massi et al.\ \cite{massi:06}; IRS22, this
work) images to better show the differences between the two bands. 
Clearly, the diffuse emission in the
$L$ band is not as prominent as in the $K_{\rm s}$ band. The spatial resolution in $L$ also
appears better than in $K_{s}$. Furthermore, the brightest $K_{s}$ sources appear not
saturated in $L$. All these facts enable us to resolve sources in $L$ that are 
not resolved in $K_{s}$ or lie within intense diffuse emission. Note also
that we have better spatial resolution, and none of the sources is saturated, with
respect to Spitzer data.

A clear example is the central bright source found by 
Massi et al.\ (\cite{massi:03}) in IRS16 and barely
resolved in two stars, their \# 90 and \# 196. Our $L$ image separates
the two objects and further shows a fainter star in between
them. The two $L$ stars without a $K_{\rm s}$ counterpart both
lie towards this group of stars, as well. 
In the $L$ image of IRS22, source \# 111 of Massi et al.\ 
(\cite{massi:03}) appears actually composed of two close-by stars. 
Furthermore, four of the $L$ stars without a $K_{\rm s}$ counterpart lie towards the
compact group of sources around \# 111.

%
%   IMMAGINE L: IRS16                            One column figure
%----------------------------------------------------------- 
%
%______________________________________________________________

%
%   IMMAGINE L: IRS17                            One column figure
%----------------------------------------------------------- 
%
%______________________________________________________________

%
%   IMMAGINE L: IRS18                            One column figure
%----------------------------------------------------------- 
%
%______________________________________________________________

%
%   IMMAGINE L: IRS20                            One column figure
%----------------------------------------------------------- 
%
%______________________________________________________________

%
%   IMMAGINE L: IRS21                            One column figure
%----------------------------------------------------------- 
%
%______________________________________________________________

%
%   IMMAGINE L: IRS22                            One column figure
%----------------------------------------------------------- 
%
%______________________________________________________________

\subsection{The cluster stellar population}
\label{star:pop}

\begin{figure*}
   \centering
   \includegraphics[width=15cm]{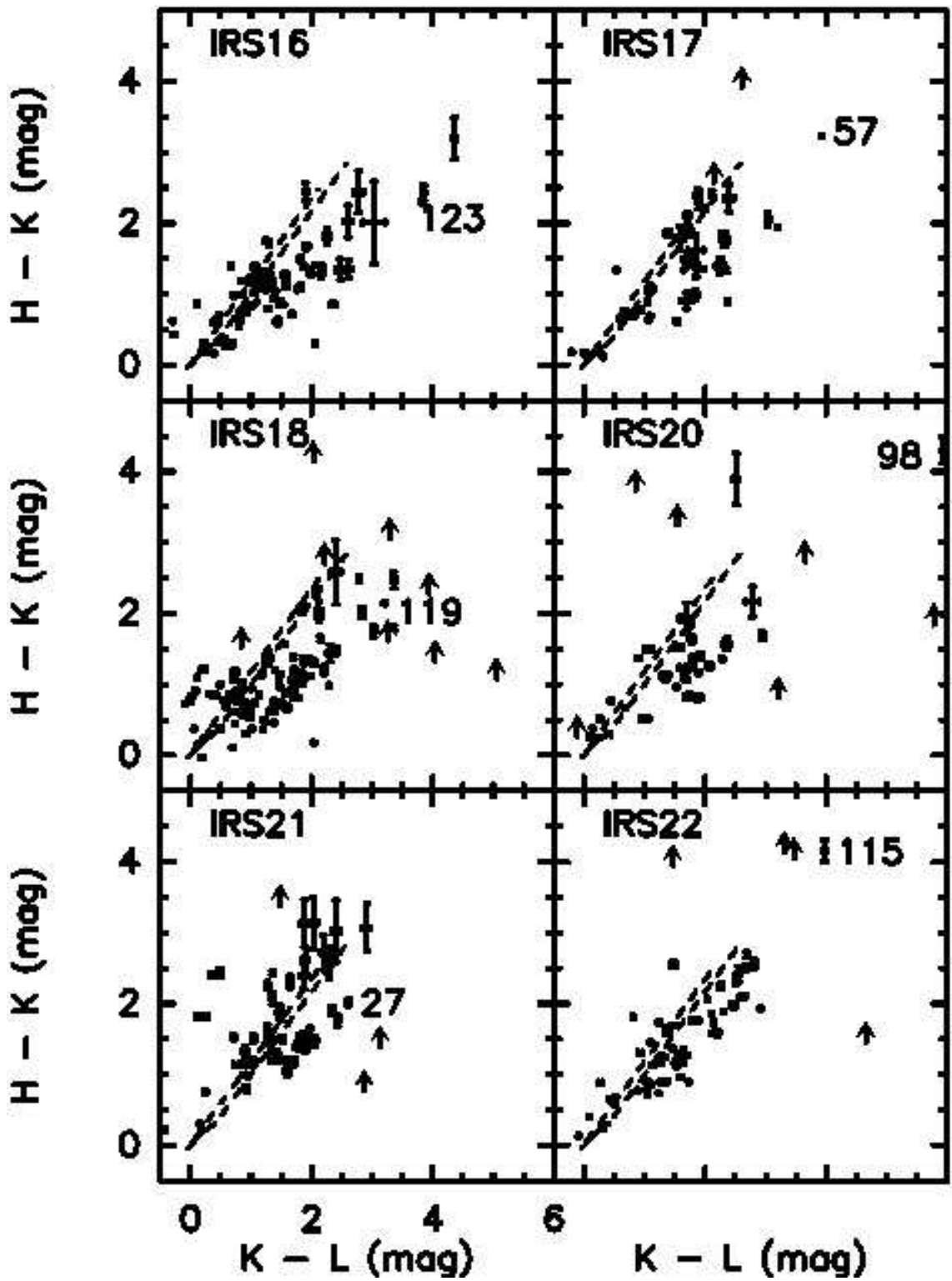}
   \caption{$H - K_{\rm s}$ vs.\ $K_{\rm s} - L$ diagrams for all sources with $L < 14.25$
     (the completeness limit for half of the clusters) 
     in each field. Arrows mark lower
      limits in $H - K_{\rm s}$. The locus of main sequence stars
      (Bessell \& Brett \cite{bess:brett}) is drawn (full line)
      along with the loci of reddened B8 and M5 dwarfs up to
      $A_{V} = 40$ mag (dashed lines)
      according to the law of Rieke \& Lebofsky (\cite{rie:le}).
      A few remarkable sources discussed in
      the text are labelled using the numbers in the catalogues
      of Massi et al. \ (\cite{massi:99}, \cite{massi:03}).
              \label{tutti:col}}
    \end{figure*}

To study the stellar population of the clusters, we computed the NIR colours
of all sources with $L < 14.25$ (the estimated completeness limit in $L$
for half of the clusters, see Sect.~\ref{compl:sec}) and
plotted them in a $H - K_{\rm s}$ vs.\ $K_{\rm s} - L$ diagram. These diagrams 
are drawn in Fig.~\ref{tutti:col}
for all fields. To discriminate between sources with a NIR excess and reddened
stars, a knowledge both of the colours of unreddened main sequence stars
and of the reddening law is needed. According to the ISAAC manual, the available $L$ filter
is more similar to the $L'$ filter described in Bessell \& Brett (\cite{bess:brett}).
According to their Table II, $K - L$ and $K - L'$ can differ by up to $\sim 0.1$ mag
for main sequence M stars. Conversely, the values of $H - K$ are mostly within few $\sim 0.01$ mag 
with respect to those given by Koornneef (\cite{Koorn}). 
Hence, we adopted the colours (and $L'$) by Bessell \& Brett (\cite{bess:brett}) for the main
sequence locus drawn in Fig.~\ref{tutti:col}. However, note that the 6 fields were 
imaged in the $K_{\rm s}$
(SofI) band, that is slightly different from the $K$ band adopted by Bessell \& Brett (\cite{bess:brett})
or Koornneef (\cite{Koorn}). According to Persson et al.\ (\cite{persy}), the scatter
in the relationship between $K_{\rm s} - K$ and stellar colours is mostly due to the presence or lack
of stellar CO-band absorption that affects the $K$ filter more than the $K_{\rm s}$ filter. 
For the standards listed in their Table~2, the differences between $K$ and $K_{\rm s}$ are
at most less than few $\sim 0.01$ mag. Differences between $K$ and $K_{\rm s}$ can be
also derived by using the transformations given by Carpenter (\cite{carpy}) for 
2MASS. In the colour range spread by main sequence stars, the differences between $K_{\rm s}$
(2MASS) and $K$ (in the Koornneef or Bessell systems) are always $< 0.05$ mag. 
These should be of the same order for the SofI filter, so the main sequence colours 
given by Bessell \& Brett (\cite{bess:brett}) should be within $\sim 0.05$ mag at most
of the
actual ones in the SofI system. The adopted reddening law is that of Rieke \& Lebofsky 
(\cite{rie:le}), that Massi et al.\ (\cite{massi:06}) showed to be consistent with the
reddening in the SofI $JHK_{\rm s}$ system.

%  tutti i colori!                    Two column figure (place early!)
%______________________________________________ Gamma_1 (lg rho, lg e)

In all colour-colour diagrams of Fig.~\ref{tutti:col} it is quite evident
that part of the sources are aligned within the reddening band of the
main sequence, whereas a large fraction of sources exhibit a NIR excess
although they are spread along a direction parallel to the reddening vector,
as well. The NIR counterparts of
the IRAS sources (reidentified in 
%
% CORREZIONE 4
%
Appendix~\ref{appe}) 
%
% FINE CORREZIONE 4
%
appear usually 
located at the upper edge of this sequence
of sources with a NIR excess, or of its projection towards the upper-right corner
of the diagram (e.\ g., source \# 57 in IRS17 and source \# 98 in IRS20), 
meaning they are heavily reddened, as well. In this respect,
source \# 98 (IRS20) is by far the most extincted of all counterparts ($A_{V}
\sim 80$ mag), consistent with it being viewed through an edge-on
circumstellar disk.

There are also datapoints found above the reddening band of the main sequence
in all fields. 
We found that most of these sources are very close to other
sources in the $K_{\rm s}$ frames, so that the aperture photometry in $K_{\rm s}$ 
is contaminated 
by the light from the nearby stars and the corresponding $K_{\rm s} - L$ values
are moved leftwards, as well.

The sources within the reddening band of the main sequence exhibit extinctions
up to $A_{V}= 20 - 40$ mag. This and the high fraction of sources with a
NIR excess, indicate that the fields imaged in $L$ are dominated by the
young members of the embedded clusters. 

%  DENSITA' SUPERFICIALE! two column figure (place early!)
%______________________________________________ Gamma_1 (lg rho, lg e)
   \begin{figure*}
   \centering
   \includegraphics[angle=-90,width=16cm]{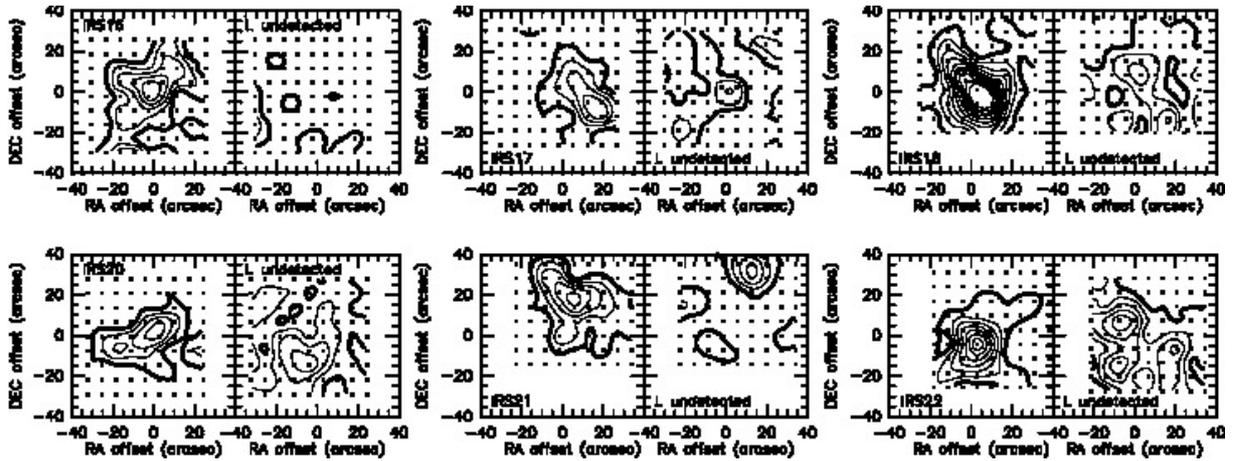}
   \caption{Maps of the surface density of stars towards the observed fields.
     In the left-hand boxes the surface density of sources with $L < 14.25$ is
     plotted, whereas in the right-hand boxes the surface density of $K_{\rm s}$ 
     sources
     without an $L$ counterpart is shown. The lowest contour (thick line)
     is 40 stars arcmin$^{-2}$, the other contours are in steps of 40 stars 
     arcmin$^{-2}$. Coordinates are offsets from the position of the
     identified counterparts of the IRAS sources.
              \label{tutte:mappe}}
    \end{figure*}

To further prove that the imaged $L$ fields are centred towards the cluster cores,
we plot the surface density of $L$ sources in Fig.~\ref{tutte:mappe} for all the fields
(left boxes). The surface density is obtained by counting all sources with $L < 14.25$ within
$14 \arcsec$-side squares, displaced by $7 \arcsec$ in right ascension and declination
from each other. 
The surface density maps are centred on the identified NIR counterparts of the
IRAS sources (see 
%
% CORREZIONE 5
%
Appendix~\ref{appe}).
%
% FINE CORREZIONE 5
%
By binning together all counts per cell towards the 6 clusters, we obtained a histogram
that is reminiscent of a Poissonian curve with 1 average and an excess of counts
on the wing. This allows us to roughly estimate a standard deviation of
20 stars arcmin$^{-2}$ towards all fields. Clearly, all cluster cores fall
within the imaged fields. Note that some of the proposed counterparts are not
located in the centre of the clusters (e.\ g., source \# 57 in IRS17).
 
%
%  MAG-COL DIAGRAM! two column figure (place early!)
%______________________________________________ Gamma_1 (lg rho, lg e)
   \begin{figure*}
   \centering
   \includegraphics[angle=-90,width=16cm]{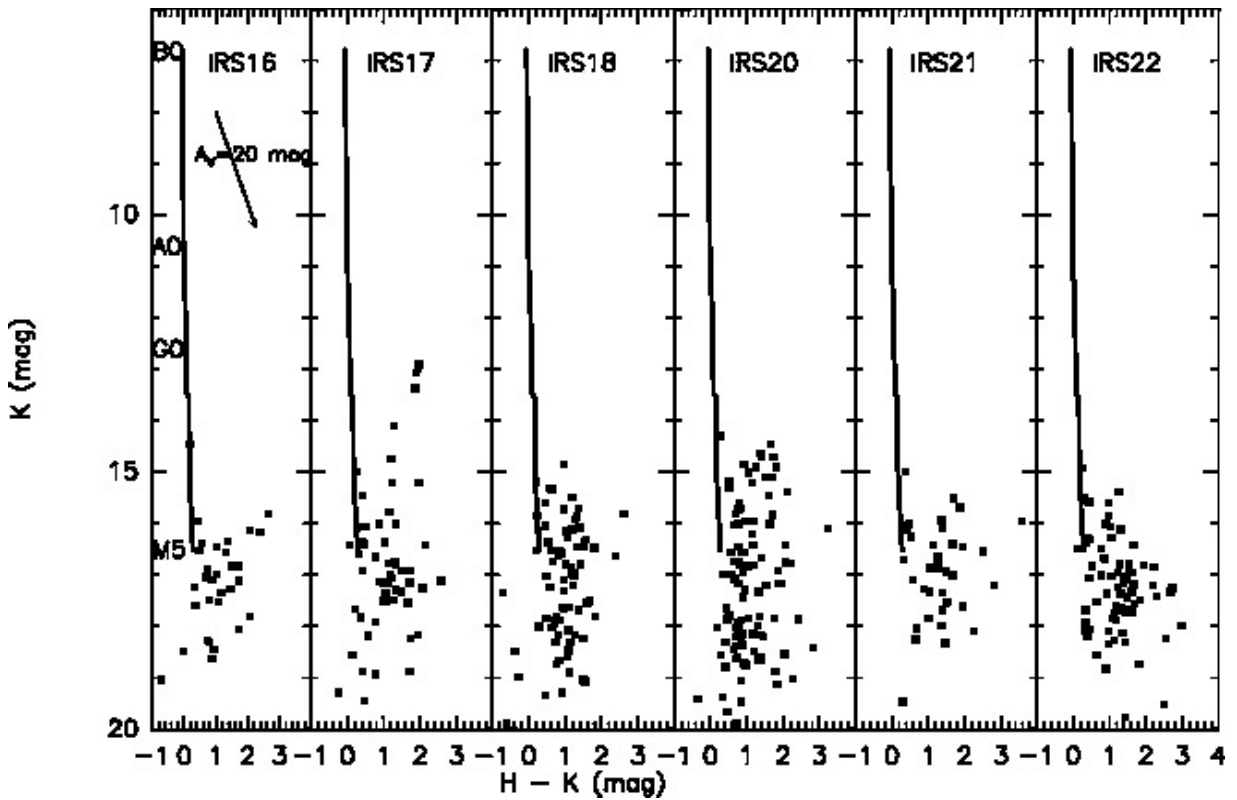}
   \caption{$K_{\rm s}$ vs.\ $H - K_{\rm s}$ diagrams for all $K_{\rm s}$ 
    sources in the observed fields
    without an $L$ counterpart. The locus of the ZAMS at $d=700$ pc is also drawn,
    adopting the colours of Koornneef (\cite{Koorn}) and the absolute magnitudes given
    by Allen (\cite{allen}). A few spectral types are labelled and a reddening vector
    $A_{V}=20$ mag is drawn, as well, according to Rieke \& Lebofsky (\cite{rie:le}). 
              \label{mag:col}}
    \end{figure*}
%______________________________________________

\subsection{The population of NIR stars without an $L$ counterpart}

In Sect.~\ref{obs:sec}, we noted the presence of $K_{\rm s}$ sources without
an $L$ counterpart in all fields and checked that they are consistent with
faint stars whose $L$ flux falls below the completeness limit.
This is confirmed by the magnitude-colour diagrams ($K_{\rm s}$ vs.\ $H - K_{\rm s}$)
shown in Fig.~\ref{mag:col} for all $K_{\rm s}$ sources in the imaged fields without an 
$L$ counterparts. All these sources are actually fainter than 
$K_{\rm s} \sim 14-15$. 

To further confirm their nature, in Fig.~\ref{tutte:mappe} we plot contours
of the surface 
density of $K_{\rm s}$ sources without an $L$ counterpart (right boxes) for
all fields, obtained
as explained above. Some of the fields exhibit a clustering of such sources 
that is not centred towards the spatial distribution centre of the $L$ sources. However,
note that the brightness of most of these $K_{\rm s}$ sources is below the completeness
limit in $K_{\rm s}$. Then, part of these faint sources may be cluster members, 
as well, and
their different spatial distribution may just reflect variations in the
detection efficiency due to 
variations in extinction and diffuse emission over the fields. 

\subsection{Contamination of the cluster members by field stars}
\label{contam}

\begin{figure}
   \centering
   \includegraphics[width=7cm]{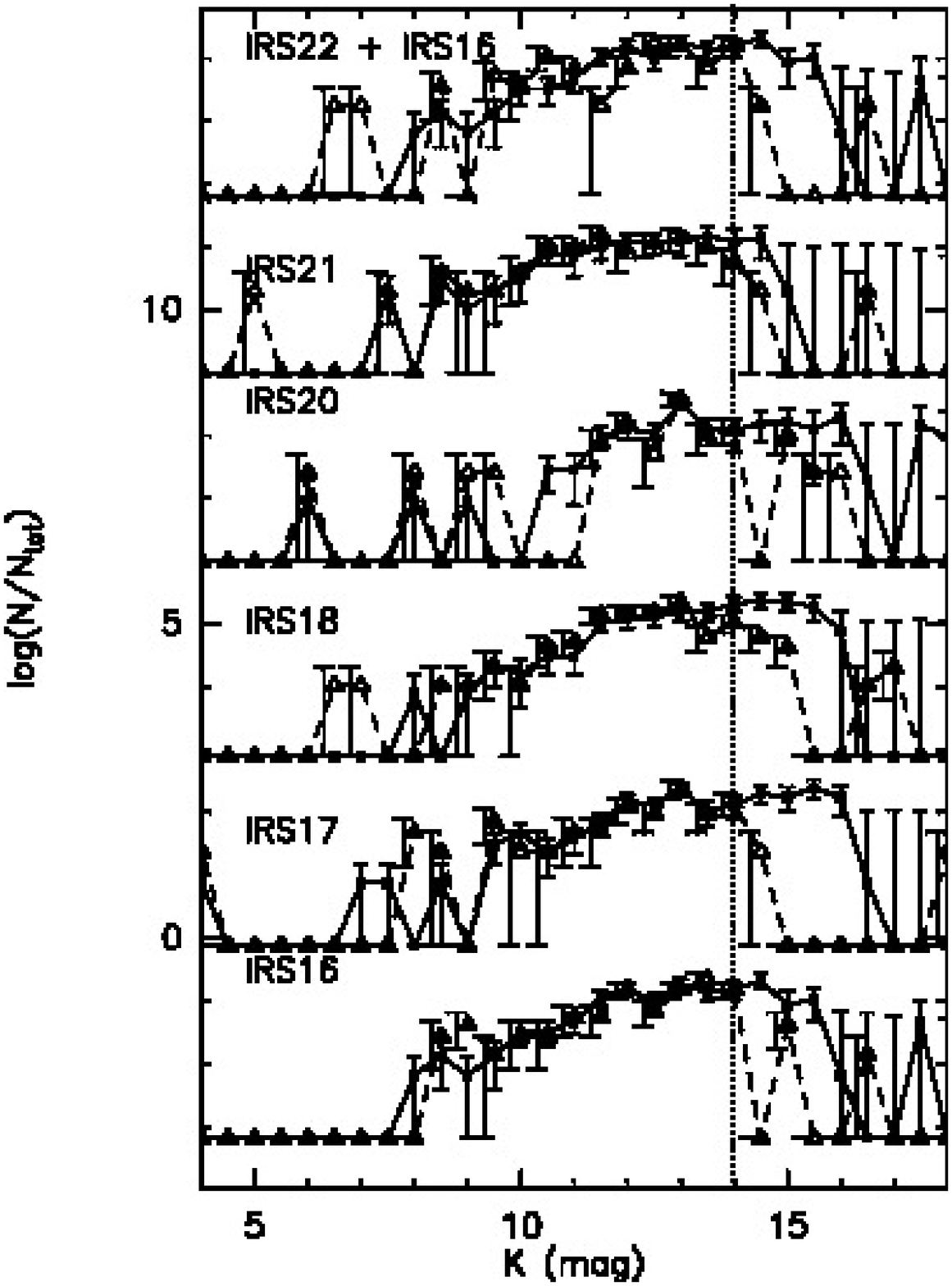}
      \caption{Dereddened $K$ luminosity functions for the 6 clusters
      obtained by using all $K_{\rm s}$ sources with an $L$ counterpart
      (open triangles, dashed lines), overlaid with the corresponding
      dereddened $K_{\rm s}$ luminosity functions by Massi et al. (\cite{massi:06};
      full squares, solid lines). Errorbars for Poissonian errors are
      also drawn (those for the $L$ sources are slightly shifted
      along the $x$-axis). The vertical dotted line marks the dereddened
      completeness limit ($K_{\rm s} \sim 14$) estimated by Massi et al.\ (\cite{massi:06})
      for their K luminosity functions.
         \label{klf:all} }
   \end{figure}

To check how much the population of $L$ sources is contaminated by
field stars, we derived the $K$ Luminosity Function (KLF) of
each cluster by using all $K_{\rm s}$ sources with an $L$ counterpart.
These KLFs were then dereddened according to the method
discussed in Massi et al.\ (\cite{massi:06}),
and compared with the ones obtained for the same clusters
by these authors after dereddening
and correction for field stars (excluded IRS22, not observed by
Massi et al.\ \cite{massi:06}). Figure~\ref{klf:all} shows that  
all pairs of KLFs are quite similar for all 5 fields 
at least down to $K_{\rm s} = 14$, i. e. the completeness limit
as estimated by Massi et al.\ (\cite{massi:06}) for their dereddened KLFs. 
Hence, the $L$ photometry samples the KLFs down to a dereddened
$K_{\rm s} = 14$, as well. Some differences are 
evident at the high luminosity ends, where the statistic is however
poor. A $\chi^{2}$ test on each pair of KLFs shows that they are
equal at a significance level $> 70$ \%.
This indicates that the $L$ sources fully sample the cluster population
and that the contamination by field stars is very low.

%
%   KLF a confronto One column figure
%----------------------------------------------------------- 
%
%______________________________________________________________

\section{Discussion}
\label{Discu:sion}

\subsection{Finding the stars with a circumstellar disk}

The selected clusters are a suitable sample to check whether the fraction of members
with circumstellar disks depends on only the cluster age or other effects have to be accounted
for. First, a number of facts suggest that all 6 clusters are roughly equally old:
\begin{enumerate}
	\item They are all associated with molecular gas (Massi et al.\ \cite{massi:99},
          \cite{massi:03}, \cite{massi:06}), see Tab.~\ref{table:fields};
	\item they are all associated with dense molecular cores (Massi et al.\ 
          \cite{massi:07}), see Tab.~\ref{table:fields};
        \item they are all associated with a FIR point source that has
          been identified (excepted for IRS16) as an intermediate-mass star progenitor
          (Massi et al.\ \cite{massi:99}, \cite{massi:03});
	\item they have roughly the same projected size (Massi et al.\ \cite{massi:99},
          \cite{massi:03});
	\item Massi et al.\ (\cite{massi:06}) showed that the best fits of their KLFs
          are obtained with pms stars of an age $\sim 1-5$ Myr;
	\item Elia et al.\ (\cite{elia:07}) found that four of the clusters are located
          within filaments of molecular gas whose origin is probably dynamical
          with an estimated age of $\sim 1$ Myr. 
\end{enumerate}

Secondly, all clusters belong to a same region (Vela-D) and are, therefore, at a same distance
and suffer a similar reddening distribution. All of them were observed in $JHK_{\rm s}L$ 
by using the same instruments and instrumental
setups. Hence, they are all affected by the same environmental and instrumental biases
and their properties are easier to compare.

Haisch et al.\ (\cite{hai:00}), Lada et al.\ (\cite{lada:00})
and Lada \& Lada (\cite{la:la}) showed that
virtually every source with a circumstellar disk exhibit a NIR excess in
a $JHKL$ diagram. Hence, the fraction of young stars with a circumstellar disk
can be derived by just counting the fraction of objects with a NIR excess
in a $JHKL$ diagram. We rather used the $HK_{\rm s}L$ diagrams of Fig.~\ref{tutti:col},
since many $HK_{\rm s}L$ sources are not detected in $J$. We adopted the main sequence
locus of Bessell \& Brett (\cite{bess:brett}) and the reddening law
of Rieke \& Lebofsky (\cite{rie:le}), as discussed in Sect.~\ref{star:pop}. 
Conservatively, given an $H - K_{\rm s}$ vs.\ $K_{\rm s} - L$ diagram
we considered as sources with a NIR excess those lying below
a line parallel to the reddening vector, passing through a point
$0.1$ mag bluer in $H - K_{\rm s}$ than an M5 star. The most optimistic mass limit
estimate (with $A_{V} \sim 40$ mag) 
corresponding to our completeness $L$ magnitude, discussed in Sect.~\ref{compl:sec},
matches a main sequence M5 star. E. g., according to Baraffe \& Chabrier (\cite{bar:cha})
the mass of an M5 V star is $\sim 0.13$ $M_{\sun}$. In principle, less extincted
low mass stars (later than M5) could be detected and counted as stars with a NIR excess.
Adopting a selection cut at a lower $L$, e. g. at $L_{\rm compl}''$, reduces
by much this kind of contamination. Furthermore, the obtained
fraction of stars exhibiting a NIR excess is likely to be a lower limit.  
In fact, whereas the $H$ and $K_{\rm s}$ images were usually taken under similar
seeing conditions, the $L$ ones are characterised by a better spatial resolution. Hence, 
$K_{\rm s} - L$ could appear bluer than it is for close-by stars, moving the corresponding 
datapoints leftwards in the diagram, i. e., increasing the apparent number of sources without
a NIR excess. Furthermore, all sources falling above the reddening band
of the main sequence have been conservatively counted as stars without a NIR excess. 

As discussed in Sect.~\ref{contam}, the contamination by field stars is expected to
be quite low. It can be estimated by noting that our photometry for stars 
with $L < 14.25$ appears to sample all sources down to a dereddened $K_{\rm s} = 14$. 
By counting all objects down to $K_{\rm s} = 14$ in the dereddened KLFs for field stars
estimated by
Massi et al.\ (\cite{massi:06}) and scaling their number to the area of the
$L$ fields, we derived 5--6 contaminant stars per field. This agrees with what found in
Sect.~\ref{contam}.

%_____________________________________________________________
%               Frazione di sorgenti con eccesso NIR!                              
%_____________________________________________________________
%
\begin{table*}
\caption{Fraction of $HK_{\rm s}L$ sources with a NIR excess. Also listed the
        fraction derived by selecting only $HK_{\rm s}L$ sources whose $K_{\rm s}$,
        after correction for reddening as explained in Sect.~\ref{frac:def}
        ($K_{\rm dered}$), is less than given thresholds
        (the selection criteria on $L$ and $K_{\rm dered}$ are indicated
        in the first line). 
\label{frac:star}}      % is used to refer this table in the text
\centering                          % used for centering table
\begin{tabular}{c c c c c c c c}        % centered columns (4 columns)
\hline\hline                 % inserts double horizontal lines
      & \multicolumn{2}{c}{$L < L_{\rm compl}''$} & 
            \multicolumn{2}{c}{$L < L_{\rm compl}$} &
             \multicolumn{2}{c}{$L < L_{\rm compl}$ and $K_{\rm dered} < 14$} & 
             \multicolumn{1}{c}{$L < L_{\rm compl}$ and $K_{\rm dered} < 11$} \\
Field & Fraction & Fraction & Fraction & Fraction &
        Fraction & Fraction & Fraction \\    % table heading
name  & corrected & uncorrected & corrected  & uncorrected  &
       corrected & uncorrected & uncorrected \\
\hline                        % inserts single horizontal line
IRS16 & $0.74 \pm 0.18$ & $0.65 \pm 0.16$ & $0.65 \pm 0.13$ & $0.60 \pm 0.11$ & 
         $0.60 \pm 0.12$ & $0.56 \pm 0.11$ & $0.46 \pm 0.25$ \\ 
IRS17 & $0.93 \pm 0.30$ & $0.65 \pm 0.23$ & $0.66 \pm 0.18$ & $0.57 \pm 0.15$ & 
         $0.59 \pm 0.16$ & $0.50 \pm 0.14$ & $0.56 \pm 0.31$ \\ 
IRS18 & $0.73 \pm 0.13$ & $0.68 \pm 0.13$ & $0.66 \pm 0.11$ & $0.63 \pm 0.10$ & 
         $0.62 \pm 0.11$ & $0.59 \pm 0.10$ & $0.43 \pm 0.21$ \\ 
IRS20 & $0.91 \pm 0.28$ & $0.68 \pm 0.23$ & $0.73 \pm 0.17$ & $0.63 \pm 0.15$ & 
         $0.66 \pm 0.17$ & $0.56 \pm 0.15$ & $0.25 \pm 0.28$ \\ 
IRS21 & $0.69 \pm 0.22$ & $0.52 \pm 0.19$ & $0.49 \pm 0.12$ & $0.43 \pm 0.11$ & 
         $0.46 \pm 0.11$ & $0.41 \pm 0.11$ & $0.29 \pm 0.16$ \\ 
IRS22 & $0.87 \pm 0.24$ & $0.66 \pm 0.20$ & $0.69 \pm 0.16$ & $0.62 \pm 0.13$ & 
         $0.67 \pm 0.15$ & $0.60 \pm 0.14$ & $0.50 \pm 0.22$ \\ 
\hline                                   %inserts single line
\end{tabular}
\end{table*}
%
%_____________________________________________________________

The fractions of stars with a NIR excess per cluster are 
listed in Tab.~\ref{frac:star}, both
with and without correction for field stars. The quoted uncertainties are 
obtained by propagating the 
Poissonian errors. We derived the fractions both including all stars down to  
$L_{\rm compl}$ and including only stars down to $L_{\rm compl}''$, to test the effect of
both the different completeness limits and the possible contamination of the NIR
excess region from 
very low mass stars. Clearly, we obtain slightly higher
fractions for the lower completeness magnitude. If these are also corrected for 
field star contamination by using the same numbers as above, 
they become even higher. But the highest values are found
for the clusters with less sources below $L_{\rm compl}''$ (IRS17 and IRS20) and
the contamination from field stars is now overestimated, as well. Hence,
the differences for each cluster remain within the errors.

\subsection{Sources of bias in counting stars with a circumstellar disk}  
\label{sec:bias}

There is a major effect that has to be taken into account when deriving
the fraction of stars with a NIR excess using cuts in magnitude. Because 
of extinction, a cut in magnitude implies that the volume in which
less massive stars are detected decreases with decreasing mass. This
would be even worse if the age spread were not negligible. If stars
of the same mass with and without a disk were detected within the same volume,
even if this changed with mass
there would be no problems in deriving a meaningful ratio. But,
because of the NIR excess, stars with a circumstellar disk can be detected
deeper in the cloud with respect to stars of the same mass without 
a circumstellar disk. So, the number of stars with a NIR excess is bound
to overestimate the number of stars with a circumstellar disk used
in computing the ratio.  
One can impose a raw limit in mass by selecting only sources with a 
dereddened $K_{\rm s}$ magnitude (see Sect.~\ref{contam}), $K_{\rm dered}$,
less than a limiting value. 
We computed new fractions by selecting only sources with 
$L < L_{\rm compl}$ and $K_{\rm dered} < 14$; they are
listed in Tab.~\ref{frac:star}, both with and without correction
for field star contamination. The derived values appear to be
slightly less than the other ones, although within the errors.
Because the NIR excess is lower in $K_{\rm s}$ than in $L$, if it is small
on average then the degree of contamination is much reduced.

We can take advantage of the fact that we observed similar clusters
belonging to the same region and with the same instrumental setups,
and safely assume that the fraction $\epsilon$ of missed stars without
a disk is the same for all clusters. If $n_{2}$ is the measured number
of stars without a NIR excess, the actual number of stars without
a NIR excess will then be $n_{2}/(1 - \epsilon)$.
It is easy to see that the actual fraction $f$ is given by
$f_{m} \times (1 - \epsilon)/(1 - f_{m} \epsilon)$, were $f_{m}$
is the measured fraction. Therefore, being $f_{m}$ roughly the same,
all derived fractions have to be corrected by a same factor.
I. e., their comparison is meaningful. And even with a pessimistic
$\epsilon = 0.3$, $f_{m}$ (given in Tab.~\ref{frac:star})
should be decreased by only $0.85-0.9$ to obtain $f$.

Since the $JHK_{\rm s}$ and $L$ observations have been carried out relatively
far in time, source variability also could affect the derived
fractions. This is discussed by Lada et al.\ (\cite{lada:00}), who 
conclude that the fraction of stars falling in the excess region only
due to variability is at most negligible in the Trapezium cluster.
Again, we expect that it affects all 6 clusters the same way anyhow.
 
\subsection{Fraction of stars with a circumstellar disk in the clusters}
\label{frac:def}

Remarkably, 5 out of 6 clusters show the same fraction of 
members with a NIR excess, within the errors. IRS21 exhibits a lower fraction 
although still within the errors. According to Massi et al.\ (\cite{massi:06}),
the $K_{\rm s}$ photometry of IRS21 has a $K_{\rm s}$ completeness magnitude 1 mag lower than 
that of the other clusters. A look at the colour-colour diagram of IRS21 shows 8
datapoints with $K_{\rm s} - L > 2$ and large photometric errors in $H - K_{\rm s}$, lying 
above the adopted boundary between stars without and stars with a NIR excess. 
Objects with such a large
value of $K_{\rm s} - L$ (i. e., $ > 2$) are usually embedded protostars (see, e.\ g.,  
Haisch et al.\ \cite{hai:00}, Lada et al.\ \cite{lada:00}). All these 8 
sources are quite faint in $H$ (with photometric errors $> 0.2$ mag)
and undetected in $J$, so it is likely that their $H - K_{\rm s}$ has been
overestimated. Dropping these sources from the counting increases the
fraction of members with a NIR excess up to $0.59$ (down
to $L_{\rm compl}$ and after correction for field stars),
whereas counting them as stars with a NIR excess increases the fraction
to $0.66$. In summary, there is evidence that IRS21, as well, has the same fraction of stars
with circumstellar disks as the remaining 5 clusters. 

We also derived the fraction of stars with a NIR excess
by imposing a lower $K_{\rm dered}$ limit
(i. e., $K_{\rm dered} < 11$) in order to select the highest mass stars
in a cluster. The results are listed in Tab.~\ref{frac:star}, as well.
Although the statistic is low and contamination by foreground star
could be not negligible, the fraction
of stars with a NIR excess is clearly lower. 
This would agree with disk lifetimes being
shorter for higher mass stars. However, spectroscopic observations
should be used to infer cluster membership for the brightest stars.

\subsection{Cluster age estimates}

In conclusion, our $JHK_{\rm s}L$ photometry of 6 young embedded clusters in Vela-C
shows that the clusters host the same fraction of stars with a circumstellar disk.
This reinforces the notion that all stars (at least the low-mass ones) 
have a circumstellar disk at birth that dissipates with time
(by accretion onto the central star, dynamical interactions, photodissociation 
and/or evolving in a
planetary system; for a review of the different mechanisms see Hollenbach
et al.\ \cite{holly:ppiv}). Interestingly,
IRS16, the only cluster of the sample associated with an H{\sc ii} region, is
not different in this respect. The ionisation source is an early B star
(Massi et al.\ \cite{massi:03}) and this suggests that at least not-extreme UV environments
(i. e., those not originated by O stars) do not significantly speed up
the dissipation of disks. This agrees with the estimated time scales for
disk dissipation by UV radiation from external sources. The latter is $< 10^{7}$ yrs
only for the outer region of disks ($> 10$ AU), that 
does not contribute much to the NIR excess, and in Trapezium-like clusters,
as shown by Hollenbach et al.\ (\cite{holly:ppiv}). 

Using the relation found by Haisch et al.\ 
(\cite{hai:01}), a fraction of $0.6$--$0.7$, as observed in the 6 clusters,
translates into an age of $\sim 2$ Myr,
in agreement with what found by Massi et al.\ (\cite{massi:06}). 
Hillenbrand (\cite{hille05}) derives the fraction of stars with a circumstellar
disk in clusters from NIR colours, as well, but in a slightly different way. By comparison
with her results, we again obtain an age for the 6 clusters 
of 1--2 Myr. Interestingly,
the presence of an H{\sc ii} region ($263.619$--$0.533$) associated with IRS16
can be used to constrain the age of the clusters. 
According to Caswell \& Haynes (\cite{cas:hay}), the size of the radio emission
at 5 GHz is $1\arcmin$. This agrees with the size measured in the $K_{\rm s}$ image,
that is $\sim 2\arcmin$. Assuming a radius of $1\arcmin$, this corresponds
to $0.2$ pc at a distance of 700 pc. Massi et al.\ (\cite{massi:03}) 
found that the ionising source is probably a B0--B2 V star. The Str\"{o}mgren
radius of an H{\sc ii} region originated by a B0--B1 ZAMS star is
$r_{\rm c} = 5.17 - 0.75 \times n_{3}^{-2/3}$ pc, where $n_{3}$ is the electron density
in units of $10^{3}$ cm$^{-3}$ (Churchwell \& Walmsley \cite{ch:wa}).
Then, $263.619$--$0.533$ appears not to have reached the Str\"{o}mgren radius,
unless it expanded in a dense ($10^{4}$--$10^{5}$ cm$^{-3}$) gas now cleared away.
Even in the case of very dense gas ($10^{5}$ cm$^{-3}$), the H{\sc ii} region would
be greater than the Str\"{o}mgren sphere only for an ionising star later than 
B1 V. Before reaching the Str\"{o}mgren radius, an H{\sc ii} region expands
at a speed exceeding the sound speed in an ionised medium, i.\ e.
$> 10$ km s$^{-1}$ (see e. g. Dyson \& Williams \cite{dyson}). Hence, $263.619$--$0.533$
would have expanded to its current radius in $< 2 \times 10^{6}$ yrs.
Even in the case of a B1 star and a dense ($10^{5}$ cm$^{-3}$) medium, the dynamical
time of the H{\sc ii} region would be short (in comparison to the time
needed to reach the Str\"{o}mgren radius), i. e. $4 \times 10^{4}$ yrs 
(Dyson \& Williams \cite{dyson}). Then, the dynamical age of the H{\sc ii} region is anyway
fully consistent with the inferred cluster age.

Given that the age of the clusters is constrained at $\sim 10^{6}$ yr, it would be 
noteworthy to find out which is the physical process that can disperse 
the circumstellar disks around the sources in such a time scale. 
As for low mass stars, of the mechanisms discussed by 
Hollenbach et al.\ (\cite{holly:ppiv}) only viscous accretion onto the central
stars seems to be efficient in dispersing the inner disks so rapidly.
Stellar encounters are irrelevant for small clusters, the time scale being at
least $\sim 10^{8}$ yrs according to Hollenbach et al.\ (\cite{holly:ppiv})
for the star densities measured by Massi et al.\ (\cite{massi:06}). This is
confirmed, e. g., by the simulations of Adams et al.\ (\cite{ada}).  

\section{Conclusions}

We have carried out imaging in the $L$ band ($3.78$ $\mu$m) with ISAAC
at the VLT, of 6 young embedded
star clusters associated with cloud D of the Vela Molecular Ridge. These are
in a similar evolutionary stage and, therefore, of the same age. The $L$
photometry is complemented with $HK_{\rm s}$ photometry from SofI images.
We used $H - K_{\rm s}$ vs.\ $K_{\rm s} - L$ diagrams to derive the fraction
of cluster members with a circumstellar disk. 
Our $L$ images have better spatial resolution and no saturation problems 
in comparison to Spitzer/IRAC images. The main sources of bias and errors are 
carefully discussed. The main points of the present study are summarised as follows: 
   \begin{enumerate}
      \item We revised the NIR infrared sources that had been
        associated with the IRAS point sources in each field.
      \item By selecting $K_{\rm s}$ sources with $L$ counterparts,
        we constructed the cluster KLFs. Once dereddened by following 
        the method outlined by Massi et al.\ (\cite{massi:06}),
        we showed that these are statistically consistent with those obtained
        by Massi et al.\ (\cite{massi:06}). This demonstrates that the
        $L$ images provide significant samples of the cluster populations
        and allowed us to estimate the degree of
        contamination from field stars. 
      \item Using different selection criteria (also based on
        the dereddened KLFs), we estimated a fraction of
        stars with a circumstellar disks $\sim 0.6-0.7$ that
        appears to be the same for all clusters, within errors.
        This shows that the clusters host the same fraction of stars with a
        circumstellar disk. I.\ e., the latter depends only on
        cluster age.  
      \item We showed that the impact of the main sources of bias
        is much reduced when observing clusters belonging to a
        same molecular cloud with the same instruments and instrumental
        setups, as is the present case.
      \item The derived fraction of stars with a circumstellar disk
        constrain the age of the 6 clusters to $1-2$ Myr, refining 
        previous age estimates. 
      \item The fraction of stars with a circumstellar disk appears
        to be lower for the most massive stars. This
        would agree with shorter disk lifetimes in massive stars,
        although the results could be biased by a poor statistic.
   \end{enumerate}

\begin{acknowledgements}
LV is supported by the Basal Center for Astrophysics and Associated
Technologies PFB--06 and by the Fondecyt project n.\ 1095187.
\end{acknowledgements}

\Online

\begin{appendix} 
\section{Individual sources}
\label{appe}
\begin{figure*}
   \centering
   \includegraphics[angle=-90,width=16cm]{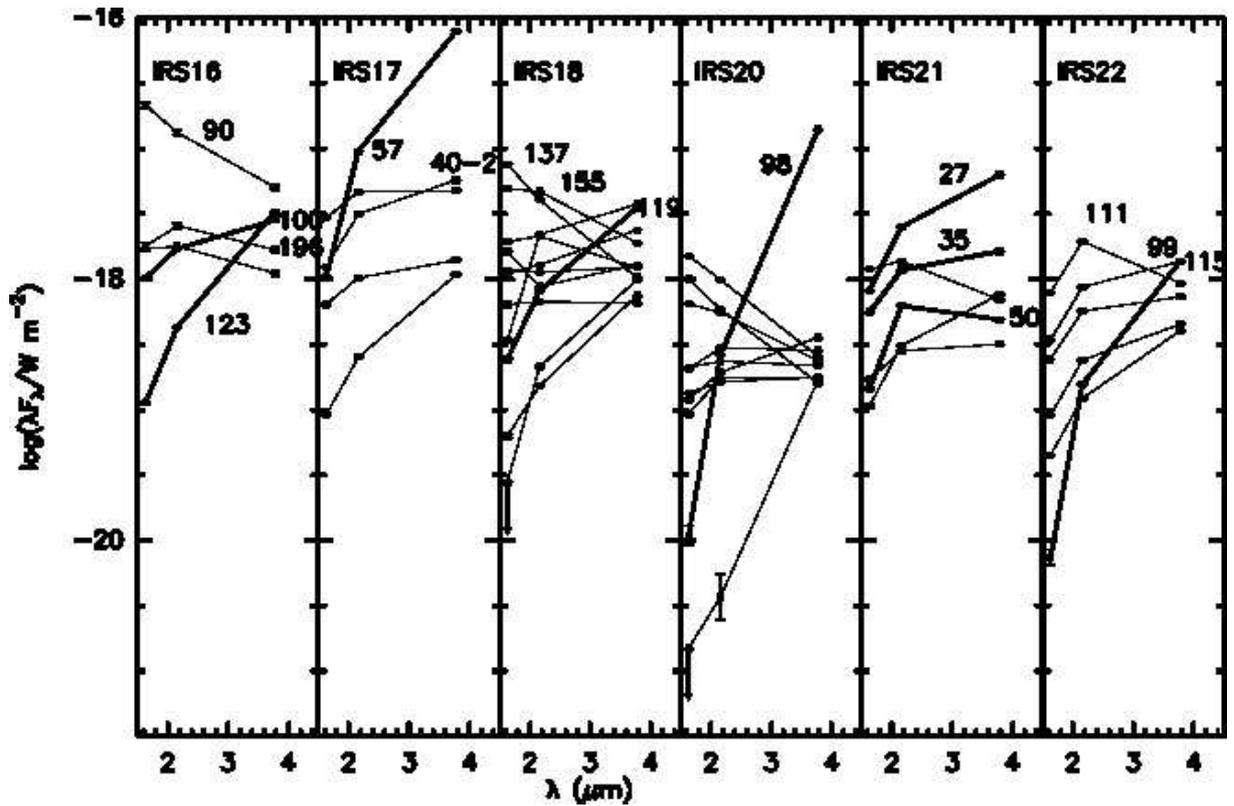}
   \caption{Spectral energy distributions for the brightest
     $L$ sources found in all 6 fields. The field name is
     on top of each box. The steeply rising SEDs
     are marked by thick lines. The most remarkable sources 
     are labelled according to the catalogue numbers by Massi et al.\
     (\cite{massi:99}, \cite{massi:03}).   
              \label{tutte:sed}}
    \end{figure*}

The observed fields were originally selected by Liseau et al.\ (\cite{liseau}),
as sites of IRAS point sources with red colours. Massi et al.\ (\cite{massi:99}, \cite{massi:03}) 
found that the fields are associated with young
embedded clusters and that all IRAS uncertainty ellipses lie towards the centres
of each cluster. These authors suggest that most of the FIR luminosity from
each IRAS point source is contributed 
by  protostars whose NIR counterparts lie within (or very near to) the uncertainty ellipse
and become brighter and brighter with increasing wavelengths. Hence, they 
should become quite evident in the $L$ band. 
To check this, the NIR SEDs of the brightest $L$ sources in each
imaged field are plotted in Fig.~\ref{tutte:sed}. Clearly, all plots (one per field)
exhibit at least one source with a steep SED whose flux increases with increasing wavelengths
and dominates over the other objects
in $L$. The $HK_{\rm s}L$ magnitudes were converted into fluxes
following M\'{e}gessier (\cite{megess}). A brief description of the 6 fields follows.
Each field will be identified by using the IRS notation of Liseau et al.\ (\cite{liseau}).

IRS16 is different from the other 5 IRAS sources in the present sample, in that 
it was not considered as a `bona fide' Class I source by Liseau et al.\
(\cite{liseau}). In fact, it is the only one associated with an H{\sc ii} region,
i.e. $263.619$--$0.533$ (Caswell \& Haynes \cite{cas:hay}). 
Massi et al.\ (\cite{massi:03}) show that the IRAS uncertainty ellipse
is roughly centred on a B0--B2 star (their \# 90, see also
Fig.~\ref{irs16_field}). Massi et al.\ (\cite{massi:07}) found that 
the IRAS uncertainty ellipse lies towards a minimum in 1.2-mm continuum emission,
in the middle of three prominent peaks (their MMS1, MMS2 and MMS3).  
It is possible that the FIR emission is originated by a UCH{\sc ii} region close
to the central B star. The source exhibiting the steepest rising SED 
in Fig.~\ref{tutte:sed}, is \# 123 of
Massi et al.\ (\cite{massi:03}), but it lies outside the uncertainty ellipse.
Another NIR source with a rising SED is \# 100
of Massi et al.\ (\cite{massi:03}), that lies within the IRAS ellipse
(see Fig.~\ref{irs16_field}). Both were already identified in $JHK$ by
Massi et al.\ (\cite{massi:03}).

The NIR counterpart of IRS17 is clearly the one already proposed by
Massi et al.\ (\cite{massi:99}), i. e. their \# 57. Giannini
et al.\ (\cite{gianni05}) showed that this NIR source lies towards a dense
molecular core found in 1.2-mm continuum emission (MMS4 of Massi et al.\
\cite{massi:07}). They also searched for
the driving source of a prominent jet, that lies towards a
group of NIR sources outside the IRAS uncertainty ellipse, on the west. These coincide with source
\# 40 of Massi et al.\ (\cite{massi:99}, see also Fig.~\ref{irs17_field}), 
resolved into more than one star
on ISAAC NIR images presented in Giannini et al.\ (\cite{gianni05}). Of these, 
\# 40--2 is bright at $L$ and exhibits a rising SEDs, as well
(Fig.~\ref{tutte:sed}). 

As for IRS18, our $L$ image again confirms that the NIR counterpart
of the IRAS source is the one already identified by Massi et al.\ 
(\cite{massi:99}), i. e. their \# 119. This exhibits
a rising SED and is one of the brightest $L$ sources in
the field (see Fig.~\ref{tutte:sed}).

No doubts that the NIR counterpart of IRS20 is the one already identified
by Massi et al.\ (\cite{massi:99}), i. e. their \# 98. It displays
a steeply rising SED and is by far the brightest $L$ source in the field
(see Fig.~\ref{tutte:sed}). Figure~\ref{irs20_field} evidences that
\# 98 lies at the centre of a bipolar nebula. Giannini et al. (\cite{gianni07})
found a bipolar jet in the H$_{2}$ 2.12-$\mu$m emission, centred at \# 98
and aligned with the nebula, suggesting that it is driven by \# 98.
This is also consistent with this source being viewed through a disk edge-on,
as suggested by Massi et al.\ (\cite{massi:03}). 

Instead, Fig.~\ref{tutte:sed} rules out that 
\# 50 of Massi et al.\ (\cite{massi:99}) is the NIR counterpart of IRS21, 
as proposed by these authors, lying within the IRAS
uncertainty ellipse. The most suited candidate appears to be \# 27
of Massi et al.\ (\cite{massi:99}), that however is located outside the IRAS
ellipse (see Fig.~\ref{irs21_field}). Source \# 35 of
Massi et al.\ (\cite{massi:99}) exhibits a rising SED, as well, but
lies outside the IRAS ellipse, as well.

Figure~\ref{tutte:sed} also suggests that the NIR counterpart of IRS22 might not be
\# 111 of Massi et al.\ (\cite{massi:03}), but its photometry
suffers from the fact that \# 111 is actually composed of two stars, unresolved in 
$H$ and $K_{\rm s}$. The most likely NIR counterpart is instead \# 115 of
Massi et al.\ (\cite{massi:03}), that exhibits a steeply rising SED and
falls within the IRAS uncertainty ellipse (see Fig.~\ref{irs22_field}).
However, we cannot rule out that one of the two sources composing \# 111 has 
a steeply rising SED. 

%  tutte le SED!                      Two column figure (place early!)
%______________________________________________ Gamma_1 (lg rho, lg e)

\end{appendix}

\end{document}